\documentstyle[11pt,newpasp,twoside,epsf]{article}
\def\ms1455     {{MS 1455.0+2232}}

\def\chandra    {{\em Chandra}\/}
\def\rxj1720    {{ RXJ1720.1+2638}\/}

\def\edcomment#1{\iffalse\marginpar{\raggedright\sl#1\/}\else\relax\fi}
\reversemarginpar

\begin{document}
\title{Chandra Observations of Cold Fronts in Cluster of Galaxies}
\author{P.\ Mazzotta, M. Markevitch,  
A. Vikhlinin, and W.R. Forman} 
\affil{Harvard-Smithsonian Center for Astrophysics, 60 Garden St.,
Cambridge, MA 02138; mazzotta@cfa.harvard.edu}

\begin{abstract} 

High-resolution \chandra\ images of several clusters of galaxies reveal
sharp, edge-like discontinuities in their gas density. The gas temperature
is higher in front of the edge where the density is low, corresponding to
approximately continuous pressure across the edge. This new phenomenon was
called ``cold fronts'', to contrast it to shock fronts that should look
similar in X-ray images but where the temperature should jump in the
opposite direction. The first cold fronts were discovered in merging
clusters, where they appear to delineate the boundaries of dense cool
subcluster remnants moving through and being stripped by the surrounding
shock-heated gas. Later, \chandra\ revealed cold fronts in the central
regions of several apparently relaxed clusters. To explain the gas bulk
motion in these clusters, we propose either a peculiar cluster formation
history that resulted in an oscillating core, or gas sloshing (without the
involvement of the underlying dark matter peak) caused by past subcluster
infall or central AGN activity. We review these observations and discuss
their implications for the X-ray cluster mass estimates.

\end{abstract}

\section{Introduction}

In the hierarchical cosmological scenario, clusters of galaxies grow through
gravitational infall and merging of smaller groups and clusters 
(e.g. White \& Rees 1978). During a merger, a significant fraction 
of the kinetic energy of the 
colliding subclusters dissipates in the intracluster medium. 
If unperturbed by further major mergers 
for few Gyr the cluster  eventually 
relaxes and the intracluster medium reaches hydrostatic equilibrium.
This status is the basic condition needed to derive the total
cluster mass distribution from X-ray observations 
(e.g. Bahcall\& Sarazin 1997).
For years clusters with azimuthally symmetric X-ray morphology and 
with X-ray brightness strongly picked on a cD galaxy have been thought 
to be the prototype of  relaxed clusters and, consequently, 
the best candidates for 
cluster mass measurements. Many of these clusters, however, 
show systematic discrepancy between the X-ray and the
 independent strong lensing derived masses inside the cluster core
(see e.g. Miralda-Escud\'e, \&  Babul 1995 and later works).
While this discrepancy may be the result of systematic effects such as
line of sight substructure and/or inadequate X-ray modeling,
it may also  indicate that the hydrostatic 
equilibrium in the cluster center may have not been achieved yet.  

Recently, \chandra ~ discovered a new phenomenon, sharp gas density
discontinuities or ``cold fronts'' that may form as a result of the relative
motion of a merging subclump (or just a distinct gas cloud) with respect to
the gas of the main cluster.

Surprisingly \chandra ~ discovered that a number of previously thought
relaxed clusters host cold fronts.  This finding, beside raising
important questions about the individual cluster formation history, 
indicates that the hydrostatic equilibrium in the central region of such
clusters may have not been achieved yet.  In this paper we briefly review
the observational evidences of cold fronts in both merging and quasi-relaxed
clusters and  discuss the implications for X-ray mass estimate.

\begin{figure}
\plottwo{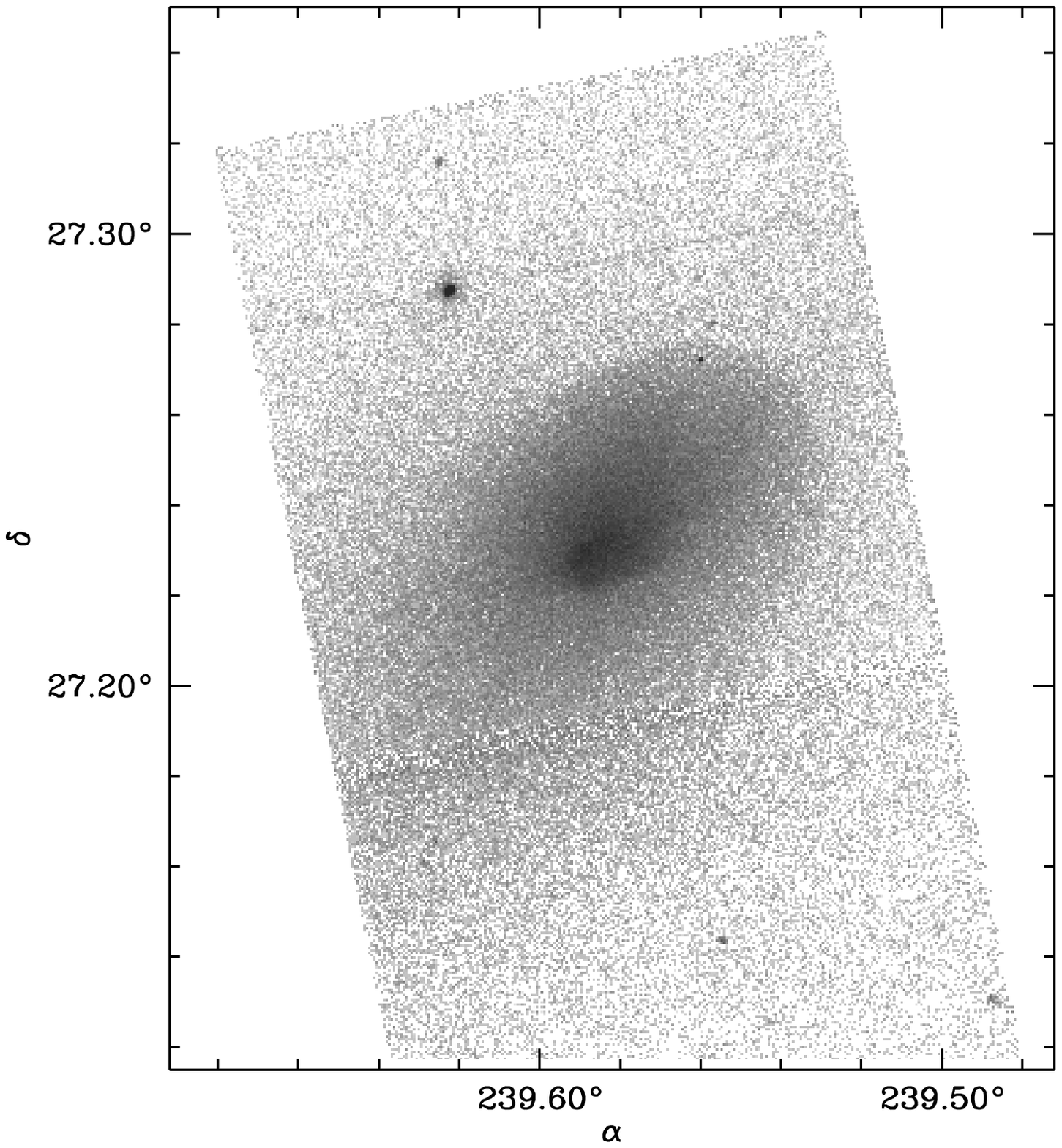}{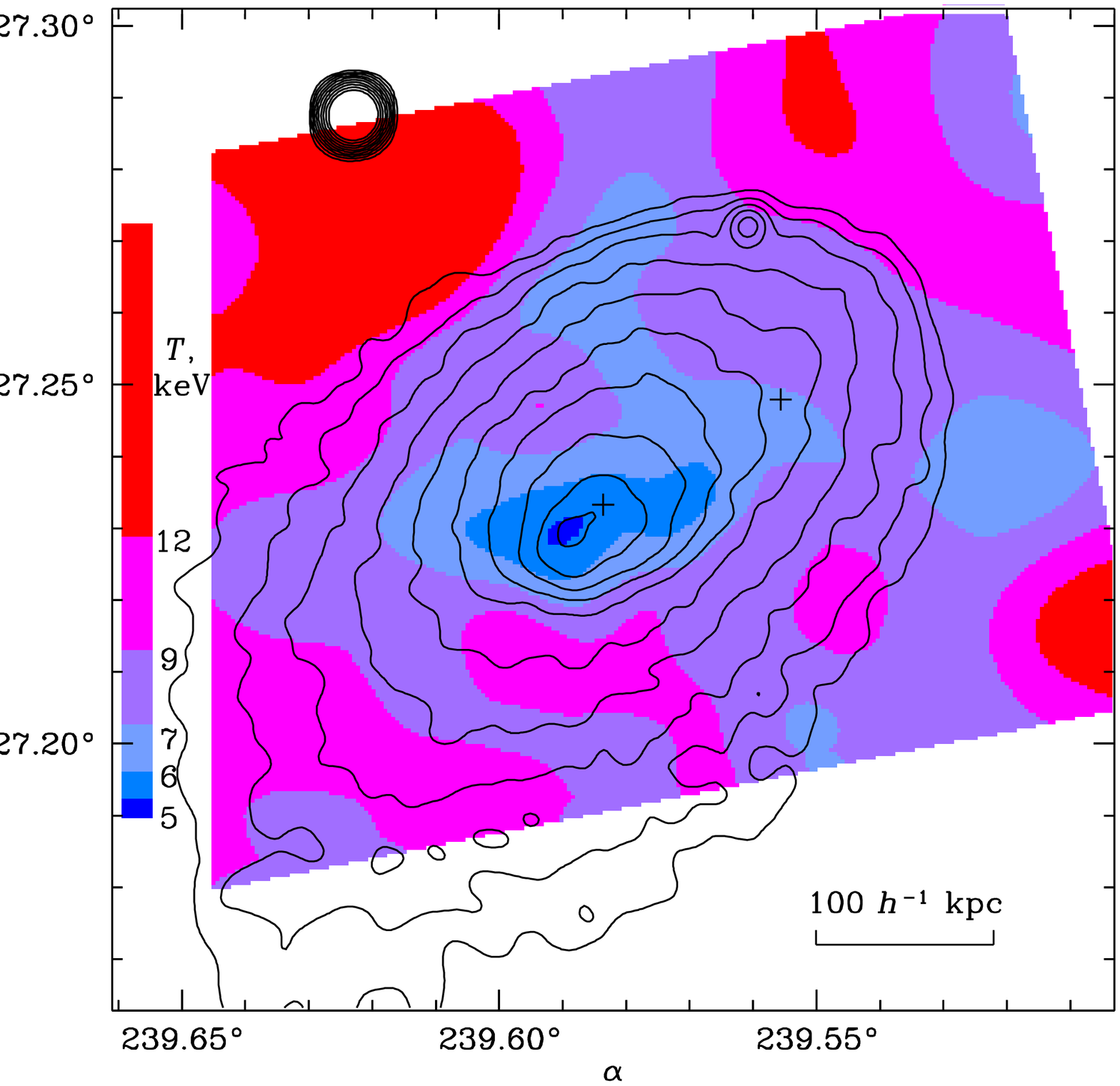}
\caption{({\em a}) Photon image of A2142 in the 0.3--10 keV band. 
Note the two sharp elliptical brightness edges northwest
and south of the cluster peak. 
({\em b}) Temperature map overlaid on the 0.3-10 keV ACIS 
brightness contours.}      
\label{fig:a2142}
\end{figure}

\section{Observation of Cold Fronts in Merger Clusters}

\begin{figure}
\plottwo{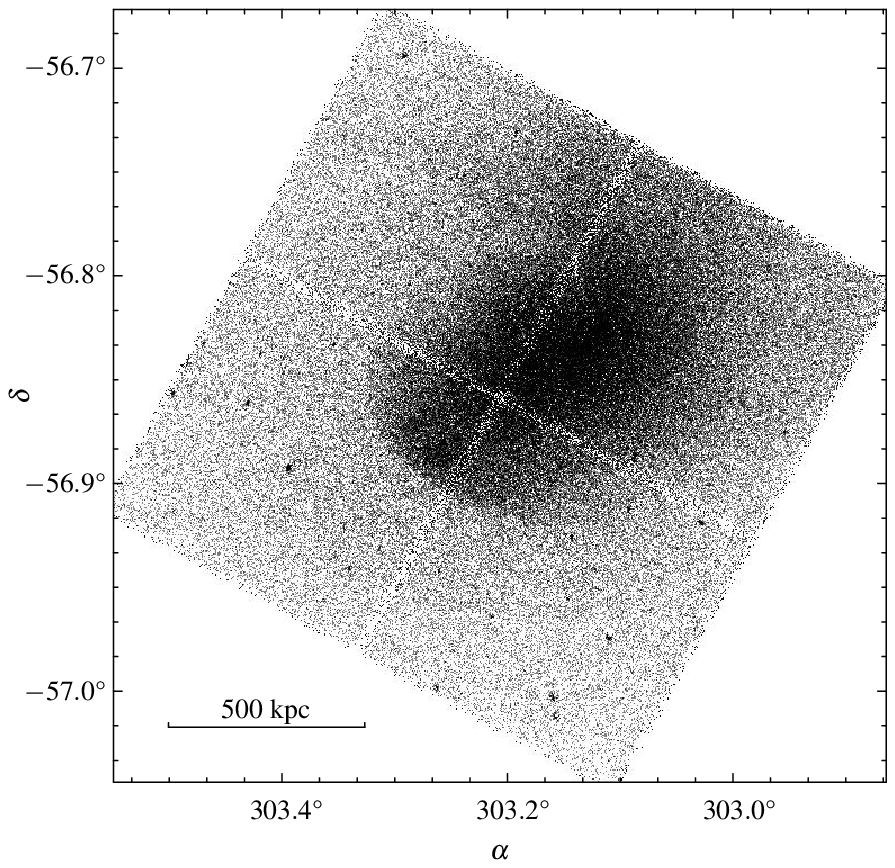}{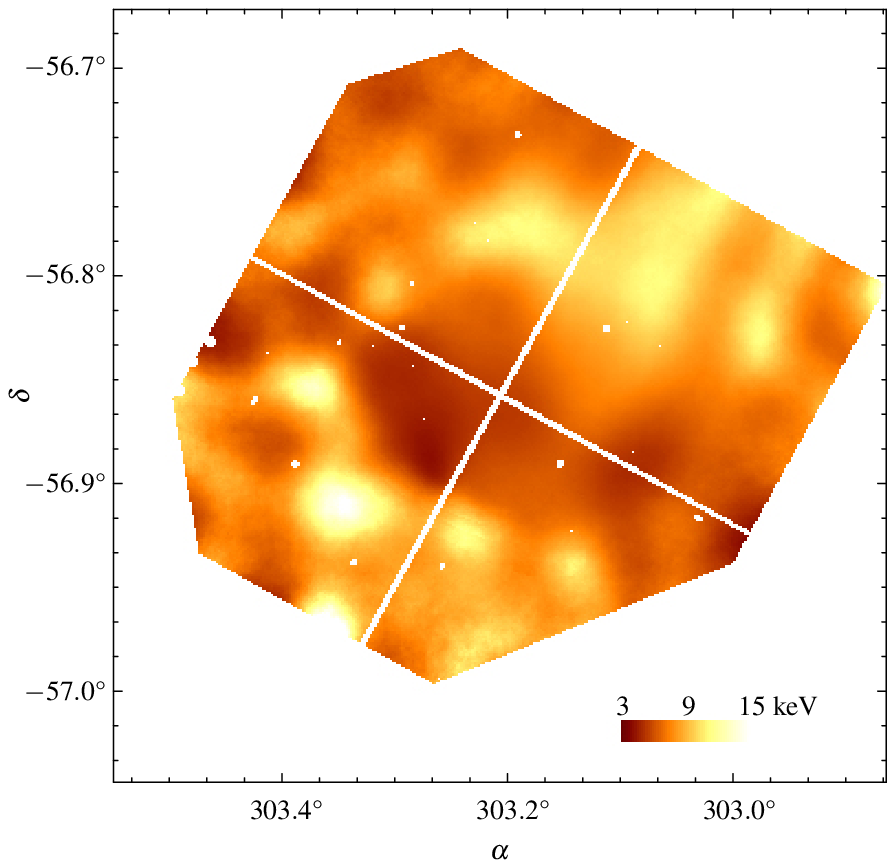}
\caption{\emph{(a)} Photon image of A3667 in the 0.5--4~keV band. 
\emph{(b)} Temperature map. The cold, $\approx 4$~keV, 
region near the center of the map coincides with the inside of the 
surface brightness edge.}
\label{fig:a3667}
\end{figure}

\begin{figure}
\plotone{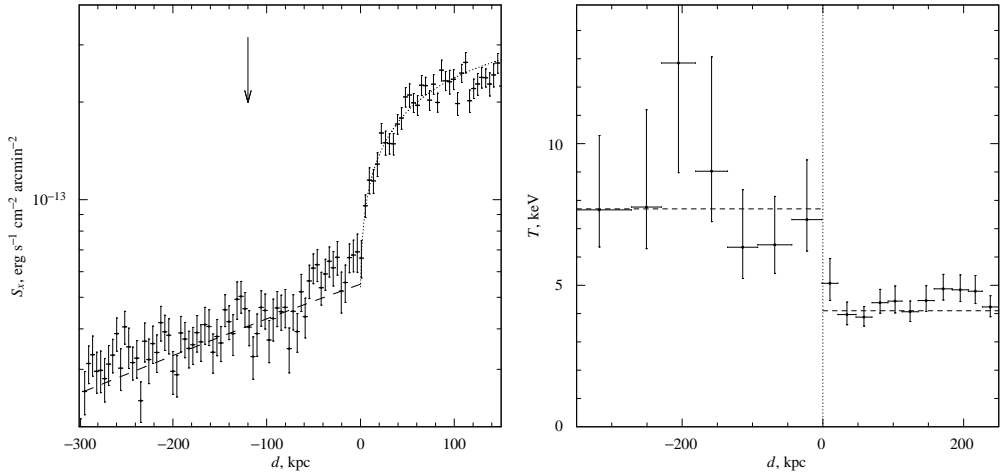}
\caption{\emph{(a)} X-ray surface brightness profile across the cold front
in A3667.
The distance is measured relative to the front position. 
\emph{(b)} Temperature profile across the front measured 
in the same sector as the surface brightness profile. 
The dashed lines show the mean gas temperature inside and outside the front.}
\label{fig:a3667_profile}
\end{figure}

The firsts cold fronts were discovered in the merging cluster A2142 
(Markevitch et al. 2000). 
How can be seen in  Fig.~\ref{fig:a2142}a,
the \chandra ~ image of this cluster reveals the presence of two sharp surface
brightness edges on opposite sides of the X-ray peak that 
indicate density profile discontinuities.
From the temperature map  it is clear that
the temperature profiles across both edges are discontinuous 
(see Fig.~\ref{fig:a2142}b).
Moreover the brightest regions across the edges are also the coldest. 
A similar surface brightness edge, with much better statistics,
 was subsequently discovered  by \chandra ~ 
in the cluster of galaxies  A3667 
(Vikhlinin, Markevitch, \& Murray 2000a; see  Fig.~\ref{fig:a3667}). 
The surface brightness and temperature profile across the edge are shown in Fig.~\ref{fig:a3667_profile}ab. These figures show that, while the brightness profile at the edge drops by a factor $\approx 5$, the temperature discontinuously
increases by a factor $\approx 2$.
It is particularly important to stress that the  
temperature variation across all these edges
goes exactly in the opposite direction to what  predicted in a shock front.
In fact, the temperature in  front of the edge is higher than that
inside where the density is higher.
For this reason these newly discovered edges have been called ``cold fronts'' 
(Vikhlinin, Markevitch, \& Murray 2000a).
One  particularly interesting aspect of the cold front is 
that the width
of the discontinuity is much less than the Coulomb free mean path. 
This finding represents a direct observation that, 
in the cold front region of the cluster,
the transport  processes in the intergalactic medium are
highly suppressed and, thus, the gasses inside and outside the edge are
thermally isolated (Ettori \& Fabian 2000; Vikhlinin et al. 2000a).
The most natural interpretation for the presence of these cold fronts is
that they represent the boundaries of cold dense gas bodies 
moving (together with their underlying dark matter concentrations)
inside  more diffuse hotter intracluster medium (Vikhlinin et al.2000a).
Indeed, from the pressure profile across the front in A3667, Vikhlinin et
al.\ have derived the subcluster velocity which corresponds to Mach number
$1\pm0.2$.
As result of this motion the magnetic field lines stretch along the front
forming an amplified parallel magnetic field that suppresses
stripping of the cool gas and stops transport processes across the boundary
(Vikhlinin et al. 2000b).

\section{Observation of Cold Fronts in ``quasi-relaxed'' Clusters}

\begin{figure*}
\plottwo{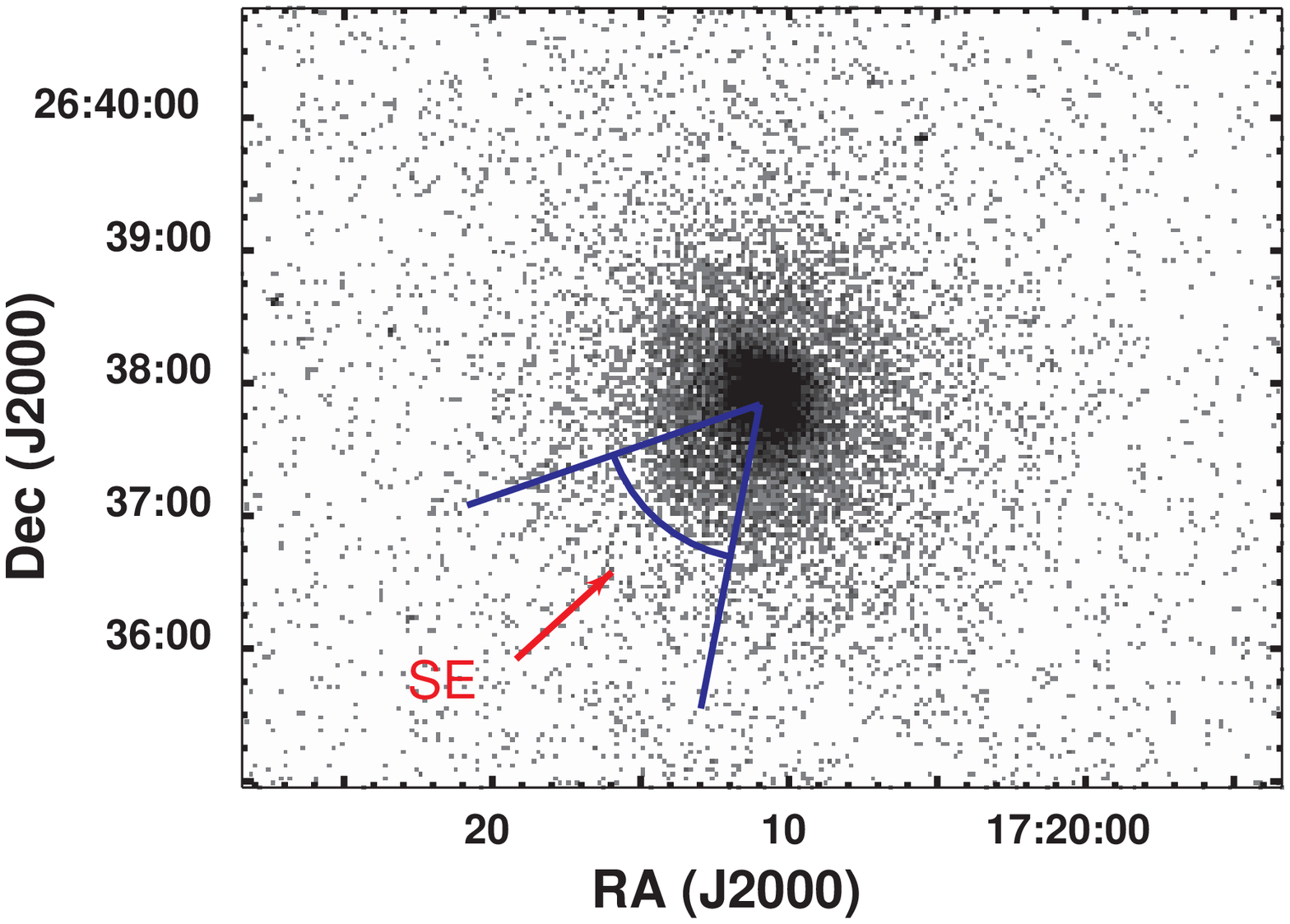}{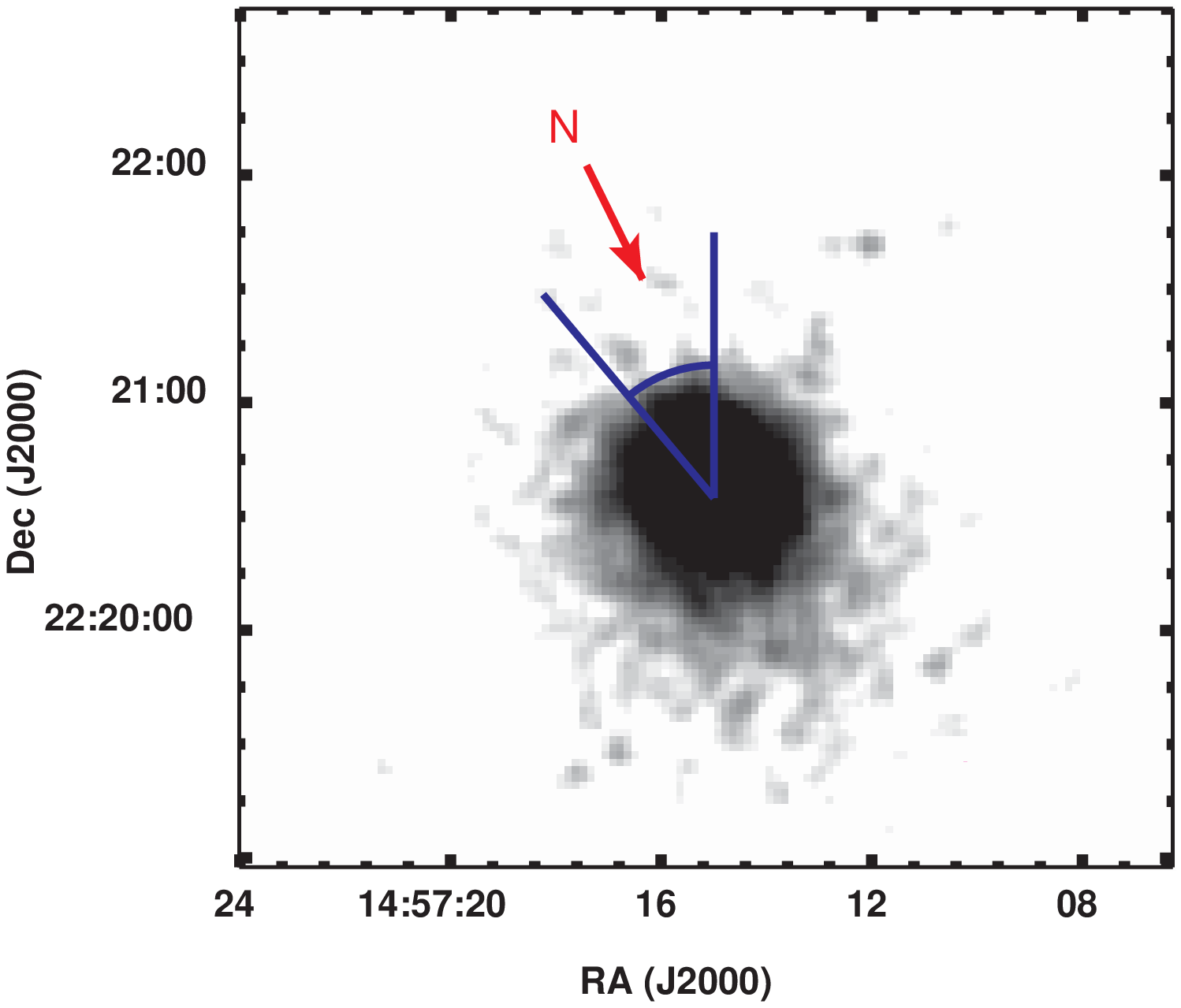}
\caption{\emph{(a)} Photon image of \rxj1720 ~ in the 
0.5-5~keV band. Note the sharp edge SE of the X-ray peak.
\emph{(b)} Photon image of \ms1455 ~ in the 
0.5-5~keV band. Note the sharp edge to the North.
}
\label{fig:round_clusters}
\end{figure*}

Beside the merger clusters, \chandra ~ discovered cold fronts
also in more round  and apparently ``relaxed clusters''. 
The first evidence of a cold front in a round clusters
was observed in \rxj1720 ~(Mazzotta et al. 2001a). 
The \chandra ~ image of this cluster shows a surface brightness 
edge with a jump  factor of $\approx 8$ at $\approx 250$~kpc 
south-east of the X-ray peak 
(see Fig.\ref{fig:round_clusters}a). 
This surface brightness edge
corresponds to a  density jump of a factor $\approx 2.8$. 
Even though the statistics are rather poor, the temperature profile across
the edge allows us to exclude at $>3\ \sigma$ significance level 
that the observed edge is a shock front and strongly indicate that it is 
a cold front.
A similar, but stronger, edge was discovered  later 
in the cluster of galaxies \ms1455 ~(Mazzotta 
et al. 2001b).
The \chandra ~ image of this clusters shows a surface brightness edge
with a jump  factor 
of $\approx 10$ at $\approx 190$~kpc to the north of the X-ray peak 
(see Fig.\ref{fig:round_clusters}a) which corresponds to a density jump 
of a factor $\approx 3$.
\chandra ~ data on a number of other relaxed clusters at low   
(e.g. A1795 at z=0.062; Markevitch, Vikhlinin, \& Mazzotta 2001) 
as well as at high redshift (e.g. 1WGAJ1226.9+3332 at z=0.89; 
Mazzotta et al. in preparation) exhibit similar fronts in their central region
indicating that such a phenomenon may be quite common in clusters.

\subsection {Cold fronts and cluster dynamic}

The presence of cold fronts indicate that these apparently relaxed clusters
host moving group-size ($80-250$~kpc) gas clouds. The gas velocity in
relaxed clusters should be low; indeed, in A1795, it is near zero
(Markevitch, Vikhlinin \& Mazzotta 2001). The upper limit on the cloud
speeds for both \rxj1720 ~ and \ms1455 ~ show that they are moving
subsonically, although the statistical accuracy for these clusters is
poor. As the shape of the surface brightness profile is a strong
function of the direction of motion (Mazzotta et al. 2001a), we know that
the central gas clouds in these two clusters are moving in a direction close
to the plane of the sky.

At the moment it is not clear why such round clusters host moving
group size gas clouds. 
For both \rxj1720 ~ and \ms1455 ~ it has been assumed
that the gas traces the dark matter so that the clouds 
are moving with their own distinct dark matter halos.
If this is the case the most natural explanation for the presence 
of a moving subclumps is that these two 
clusters are actually merging clusters.
This scenario, however appears to be unlikely as both clusters show 
no further signs of ongoing merger as, for example, 
the typical X-ray elongation in 
the merger direction and/or the displacement of the cD galaxy with respect 
to the X-ray peak. Moreover, while the speed of a mass point free falling 
from infinity into the cluster center should be $\approx 3$
 times the speed of sound,
the observed subclump speed is clearly subsonic.
In addition we notice that 
\ms1455 ~ hosts one of the most massive 
cooling flows observed ($\dot M \approx 1500 M_\odot$~yr$^{-1}$; 
Allen et al. 1996) and it is not clear if that a massive cooling 
flow (or cool central core) would have survived the merger (see e.g. Roettiger, Loken, \& Burns 1997, but Fabian \& Daines 1991).

An alternative scenario is that, as proposed by Mazzotta et al. 2001a, 
both clusters  are the 
result of the collapse of two different  perturbations 
in the primordial density field
on two different linear scales at nearly the same location in space.
As the density field evolves, both perturbations start to collapse. 
 The small scale perturbation collapses first 
and forms a central group of galaxies while the larger perturbation
continues to evolve to form 
a more extended
 cluster potential.
The central group of galaxies could have formed  slightly offset
from the center of the cluster and is now
falling into  or oscillating around the
minimum of the cluster potential well. This motion is responsible for
the observed surface brightness discontinuity.
As the initial position of the subclump lies well within  
the  main cluster, we may also  expect the velocity of the infalling 
subclump to be  subsonic.

Another possibility is that, as proposed for A1795, the dark matter
distribution is near-stationary, but the gas is decoupled from it and
sloshing in the cluster gravitational potential (Markevich et al.  2001).
Such gas bulk motion might be caused by a disturbance of the central
gravitational potential by past subcluster infall. Alternatively, in
clusters where the central AGN produces large bubbles of hot plasma (e.g.,
McNamara et al.\ 2000), such activity could supply kinetic energy to the
surrounding cool gas. This scenario seems to be supported by simulations
of cluster mergers (that did not include AGN activity) showing that in
the final stages of a merger, the gas decouples from dark matter and
sloshes in the cluster gravitational potential for a significant time before
coming to a hydrostatic equilibrium (e.g. Roettiger et al. 1997).

\section{Cold fronts and X-ray mass estimates}\label{par:mass}  

Regardless of the origin of the cold fronts in round clusters Mazzotta et
al. 2001a and Markevitch et al. 2001 showed that its presence affects the
estimate of the total mass in the inner regions of the clusters. In
particular they showed that the mass estimate derived from the cluster
sector containing the cold front and assuming hydrostatic equilibrium has
an unphysical discontinuity at the surface brightness edge. Moreover the
inferred mass inside the edge is much lower than the value derived using
sectors of the cluster containing no cold fronts (see Fig.~4 of Mazzotta et
al. 2001a and Fig.~2e of Markevitch et al. 2001).  As a result of this
effect, measurements of the total mass using the hydrostatic equation in
circular annuli would result in a underestimate of the total mass at small
radii.  Thus, such a phenomenon may explain, in part, the discrepancy
between the X-ray and the strong lensing mass determinations found in some
systems (e.g. Miralda-Escud\'e, \& Babul 1995, but see Allen et
al.\ 2001 and reference therein).

\acknowledgments
 
P.M. acknowledges an ESA fellowship and thanks the Center for
Astrophysics for its hospitality. 
P.M. thanks the local committee for the excellent organization of the 
conference, for the wonderful visit of Taroko National Park, and ... for the 
extra fruits.  
Support for this study was provided
by NASA contract NAS8-39073, grant NAG5-3064, and by the Smithsonian
Institution.

\end{document}